\documentclass[12pt,a4paper,final]{iopart}
\usepackage{graphicx}
\usepackage{iopams}
\usepackage[colorlinks=true,allcolors=blue,breaklinks=true]{hyperref}

\def\change{\color{black}}

\newcommand*\chem[1]{\ensuremath{\mathrm{#1}}}
\newcommand{\DTO}{\chem{Dy_2Ti_2O_7}}
\newcommand{\HTO}{\chem{Ho_2Ti_2O_7}}
\newcommand{\angstrom}{\textup{\AA}}

\begin{document}

\title[Anomalous out-of-equilibrium dynamics in the spin-ice material \DTO\ under...]{Anomalous out-of-equilibrium dynamics in the spin-ice material \DTO\ under moderate magnetic fields}
\author{P C Guruciaga,$^1$ L Pili,$^{2,3}$ S Boyeras,$^{2}$\footnote{Current address: Unidad de Investigaci\'on y Desarrollo de las Ingenier\'{\i}as (UIDI), Universidad Tecnol\'ogica Nacional, Facultad Regional Buenos Aires, Medrano 951 (C1179AAQ), Buenos Aires, Argentina.} D Slobinsky,$^{2,4}$ S A Grigera$^{2,3,5}$ and R A Borzi$^{2,3}$}

\address{$^1$ Centro At\'omico Bariloche, Comisi\'on Nacional de Energ\'{\i}a At\'omica (CNEA), Consejo Nacional de Investigaciones Cient\'{\i}ficas y T\'ecnicas (CONICET), Av. E. Bustillo 9500, R8402AGP San Carlos de Bariloche, R\'{\i}o Negro, Argentina}
\address{$^2$ Instituto de F\'{\i}sica de L\'{\i}quidos y Sistemas Biol\'ogicos (IFLYSIB), UNLP-CONICET, B1900BTE La Plata, Argentina}
\address{$^3$ Departamento de F\'{\i}sica, Facultad de Ciencias Exactas, Universidad Nacional de La Plata, c.c. 16 suc. 4, B1900AJL La Plata, Argentina}
\address{$^4$ Departamento de Ingenier\'{\i}a Mec\'anica, Facultad Regional La Plata, Universidad Tecnol\'ogica Nacional, Av. 60 esq. 124, 1900 La Plata, Argentina}
\address{$^5$ School of Physics and Astronomy, University of St. Andrews, St. Andrews KY16 9SS, United Kingdom}

\ead{pamela.guruciaga@cab.cnea.gov.ar}

\begin{abstract}
    {\change {We study experimentally and numerically the dynamics of the spin ice material \DTO\ in the low temperature ($T$) and moderate magnetic field ($\bi{B}$) regime ($T \in [0.1,1.7]\,\mathrm{K}$, $B \in [0,0.3]\,\mathrm{T}$). 
    Our objective is to understand the main physics shaping the out-of-equilibrium magnetisation vs. temperature curves in two different regimes. Very far from equilibrium, turning on the magnetic field after having cooled the system in zero field (ZFC) can increase the concentration of magnetic monopoles (localised thermal excitations present in these systems); this accelerates the dynamics. Similarly to electrolytes, this occurs through dissociation of bound monopole pairs. However, for spin ices the polarisation of the vacuum out of which the monopole pairs are created is a key factor shaping the magnetisation curves, with no analog. We observe a threshold field near $0.2\,\mathrm{T}$ for this fast dynamics to take place, linked to the maximum magnetic force between the attracting pairs. Surprisingly, within a regime of low temperatures and moderate fields, an extended Ohm's law can be used to describe the ZFC magnetisation curve obtained with the dipolar spin-ice model. However, in real samples the acceleration of the dynamics appears even sharper than in simulations, possibly due to the presence of avalanches. On the other hand, the effect of the field nearer equilibrium can be just the opposite to that at very low temperatures. Single crystals, as noted before for powders, abandon equilibrium at a blocking temperature $T_{\mathrm{B}}$ which increases with field. Curiously, this behaviour is present in numerical simulations even within the nearest-neighbours interactions model. Simulations and experiments show that the increasing trend in $T_{\mathrm{B}}$ is stronger for $\bi{B}\parallel [100]$. This suggests that the field plays a part in the dynamical arrest through monopole suppression, which is quite manifest for this field orientation.}}
\end{abstract}





\section{Introduction}

The dynamics of frustrated magnetic systems~\cite{moessner1998low} is usually as appealing as their thermodynamics.
This is particularly true for spin ices, frustrated magnetic materials which have been shown to remain disordered down to the lowest temperatures~\cite{Harris1997,harris1998magnetic,Henelius2019dipolar}.
{In these materials, the spin flipping processes can be associated with the creation, annihilation and propagation of local, topological excitations (\emph{magnetic monopoles}). They do so in what otherwise would be the  massively quasi-degenerate ground state  of the system (the \textit{vacuum} of monopoles)~\cite{Castelnovo2008,castelnovo2011debye}}. 
The spin dynamics is then regulated by the density of these excitations and shaped by
the structure
of the quasi-particle vacuum~\cite{castelnovo2010thermal,jaubert2011magnetic} that acts as a dynamical constraint~\cite{mostame2014tunable}. 
In this respect, applying a magnetic field has two consequences: it alters this underlying structure ---{and might even change the dimensionality of the system ~\cite{takatsu2013two}}--- and at the same time it modifies the equilibrium density of excitations. When magnetized, a system needs to transfer
its Zeeman energy to other degrees of freedom (typically, vibrational). The magnetic coupling with the crystal lattice is then another variable needed to explain the dynamical behavior of the system. On the weakly coupled limit,
magnetic deflagrations in the form of monopole avalanches accompanied by strong increases in magnetization and temperature have been found in \DTO\ and \HTO~\cite{fennell2005neutron,slobinsky2010unconventional,jackson2014dynamic,paulsen2014far}. 
On the opposite limit, in \chem{Tb_2Ti_2O_7}, a pyrochlore close to spin ice,
the spin-lattice coupling is so strong that it leads to mixed magnetoelastic excitations~\cite{Fennell2014magnetoelastic}; its dynamics has been observed to remain unfrozen down to the lowest temperatures~\cite{gardner1999cooperative,lhotel2012low}. 
Interactions also play a very important role in the dynamics of these materials. Dipolar interactions {between magnetic moments} translate into Coulomb-like forces between monopoles, which affect their abundance and mobility ~\cite{Castelnovo2008,bramwell2009measurement,matsuhira2011spin}. The presence of these long range forces is essential to describe the dynamical freezing observed in \DTO\ and \HTO\ near $0.65\,\mathrm{K}$~\cite{jaubert2009signature} at zero magnetic field. Finally, as in other systems~\cite{castellani2005spin}, quenched disorder, in the form of impurities, lattice defects or stuffing, is also expected to play a relevant role in the dynamics
 ~\cite{revell2013evidence,lago2014glassy,snyder2004quantum,sen2015topological}.

There has been a great number of experimental and theoretical studies on the dynamics of spin ices~\cite{castelnovo2011debye,castelnovo2010thermal,jaubert2011magnetic,matsuhira2011spin,jaubert2009signature,revell2013evidence,snyder2004quantum,yaraskavitch2012spin,matsuhira2001novel,snyder2001spin,snyder2003quantum,snyder2004low,ehlers2004evidence,ke2008magnetothermodynamics,ruminy2016first,liu2014frustrated,liu2018magnetic,levis2013defects}. However, excluding works on field quenches~\cite{mostame2014tunable,stoter2020extremely}, the characterisation of the dynamics of single crystals in an applied field and at low temperatures (well within the spin-ice regime), is much more scarce~\cite{takatsu2013two,slobinsky2010unconventional,paulsen2014far,takatsu2013ac,pomaranski2013absence,clancy2009revisiting,kaiser2015ac,paulsen2016experimental}, and has left a number of questions unanswered. Among others, one aspect we investigate in this work is the dependence of the blocking temperature (the temperature at which the magnetic system starts to fall out of equilibrium within the time scale of a given experiment) with magnetic field, and in particular with the field orientation with respect to the crystalline axis. 
The purpose of this work is twofold. In the first place, we characterise the in-field magnetisation dynamics of the two most important spin models used in spin ice (nearest neighbours and dipolar). Secondly, we compare these results with our experiments on~\DTO, one of the canonical spin ice materials. Our aim is not to reproduce the experimental results in detail, but rather to find the minimum ingredients needed to understand some of their most salient features.

\section{System and methods}

\subsection{Models for spin ice}

\begin{figure}
    \centering
    \includegraphics[width=0.7\linewidth]{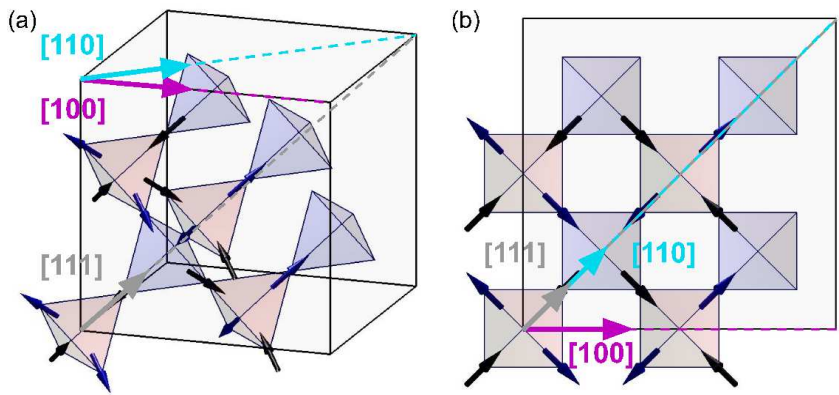}
    \caption{(a)~Conventional unit cell of the pyrochlore lattice and three directions of interest. The coloured arrows and dashed lines mark the three directions along which we have applied a magnetic field. (b)~The same configuration as seen from above.}
    \label{fig:pyroc-fields}
\end{figure}

In spin-ice materials, the magnetic moments of {size} $\mu$ can be modeled at low temperatures ($T \leq 10\,\mathrm{K}$) as Ising-like spins $\bmu_i=\mu S_i\hat{\bi{s}}_i$ occupying the vertices of a pyrochlore lattice (\fref{fig:pyroc-fields}), with $S_i=\pm 1$ and the quantisation axes $\hat{\bi{s}}_i$ {pointing} along the local $\langle 111\rangle$ directions. 
The simplest model describing these systems is the nearest-neighbour spin ice model (NNSIM), defined by
\begin{equation}\label{eq:nnsim}
    \mathcal{H}_{\mathrm{NNSIM}}= J_{\mathrm{eff}}\sum_{\langle ij\rangle}{S}_i {S}_j \, .
\end{equation}
{\change As will be discussed in the next paragraph, for real materials the constant $J_{\mathrm{eff}}$ is a combination of exchange interactions and dipolar coupling between nearest neighbours.}
The spin-ice rules (analogous to Bernal and Fowler's ice rules~\cite{Bernal1933}) are imposed by the condition $J_{\mathrm{eff}}>0$; the energy is then minimised by two spins pointing in and two pointing out of each tetrahedron. Violations of this local, divergence-free-like condition necessarily raise the energy; they will be interpreted as \emph{magnetic monopoles}~\cite{Castelnovo2008} sitting in the diamond lattice of constant {$a_{\mathrm{dia}}$} formed by the centres of the tetrahedra.  These localised excitations can be \emph{single} (3 spins in and 1 out, or vice-versa), with charge {$\pm Q=\pm 2\mu/a_{\mathrm{dia}}$}, or \emph{double} (all in, or all out), with charge $\pm 2Q$. The latter, however, are too energetic and are practically banned at temperatures such that $T/J_{\mathrm{eff}} \lesssim 1$. Within the NNSIM framework there is no effective interaction energy between monopoles~\cite{Castelnovo2008}; there are, however, entropic forces among them~\cite{henley2010coulomb}, which can be neglected to describe the spin ice materials presently known~\cite{castelnovo2011debye}.

The dipolar spin-ice model (DSIM) takes into account interactions of exchange and dipolar origin, of strengths $J$ and $D$, respectively. {Its Hamiltonian can be written as}
\begin{equation}\label{eq:dsim}
    {\mathcal H}_{\mathrm{DSIM}} = 
    J_{\mathrm{eff}}\sum_{\langle ij\rangle}{S}_i {S}_j + D\, r_{\mathrm{nn}}^3 \sum_{i>j}^{} {}^{'} \left[ \frac{\hat{\bi s}_i \cdot{\hat{\bi s}}_j}{|{\bi{r}}_{ij}|^3} 
    -\, \frac{3(\hat{\bi s}_i \cdot {\bi{r}}_{ij}) (\hat{\bi s}_j \cdot {\bi{r}}_{ij}) }{|{\bi{r}}_{ij}|^5} \right] S_i S_j \, .
\end{equation}
The angle brackets $\langle ...\rangle$ imply that only nearest neighbours are taken into account in the first sum, $r_{\mathrm{nn}}$ is {the pyrochlore lattice spacing}, and $D=\mu_0\mu^2/(4\pi r_{\mathrm{nn}}^3)$. The primed sum in the second term indicates that the nearest neighbours have been taken into account in the first term. The effective exchange constant, used also in \eref{eq:nnsim}, can be written in terms of $J$ and $D$ as $J_{\mathrm{eff}}=(J+5D)/3$. It has been shown that the inclusion of long range dipolar interaction leads to an effective Coulomb interaction between monopoles~\cite{Castelnovo2008}. Remarkably, this Hamiltonian captures not only much of the thermodynamics of spin ice materials~\cite{isakov2004magnetization,Melko04,Diep}, but also Monte Carlo simulations with the usual single-spin-flip Metropolis algorithm can describe part of their dynamical freezing~\cite{jaubert2009topological}. For \DTO~each Monte Carlo step can be associated with a characteristic spin-flip time of approximately $2.5\,\mathrm{ms}$~\cite{jaubert2011magnetic}. For simplicity, we will make use of this equivalence in order to relate characteristic times in our simulation with characteristic times measured through different experimental techniques, even though it has been shown that the attempt time can depend on temperature~\cite{revell2013evidence,takatsu2013ac} and field~\cite{takatsu2013two}.

The interaction of the spins with an external magnetic field $\bi{B}$ can be {taken into account} by adding to the corresponding Hamiltonian the Zeeman contribution
\begin{equation}\label{eq:zeeman}
    \mathcal{H}_{\mathrm{Z}}= -\mu {\bi{B}} \cdot \sum_i S_i \hat{\bi s}_i \, .
\end{equation}
In this work we {will} be interested in fields applied along three particular crystallographic directions: $[100]$, $[110]$ and $[111]$, shown in \fref{fig:pyroc-fields}(a)-(b).


\subsection{Numerical and experimental methods}\label{methods}

In order to study the dynamical behavior of \DTO, we used single-spin-flip dynamics with Metropolis algorithm in our Monte Carlo simulations. The constants in the models given by~\eref{eq:nnsim} and~\eref{eq:dsim} {\change were} set taking $J=-3.72\,\mathrm{ K}$~\cite{denHertog2000dipolar}, $\mu= 10\mu_{\mathrm{B}}$,
$r_{\mathrm{nn}}=3.5\,\angstrom${, and $a_{\mathrm{dia}}=4.3\,\angstrom$. This leads to a single monopole charge of $Q \approx  4.27 \times 10^{−13}\,\mathrm{J\,T^{-1}m^{-1}}$.} 
Long-range interactions in the DSIM were considered by means of Ewald summations~\cite{ewald1921}. We simulated cubic systems of $L^3$ conventional unit cells of the pyrochlore lattice (consisting of $16$ spins each) with periodic boundary conditions. {We used $L=3$ for the DSIM and $L=6$ for the NNSIM,}{\footnote{\change We opted in each case to run our simulations at the maximum system size which would allow us to get the results in a reasonable amount of time. We have checked that the main results described in this work hold independently of size.}} and averaged results over $100-2000$ independent runs. 
Zero-field cooling (ZFC) - field cooling (FC) protocols were simulated in the temperature range between $T_{\mathrm{min}}=100\,\mathrm{mK}$ and $T_{\mathrm{max}}=1\,\mathrm{K}$. FC magnetization curves were obtained by applying a magnetic field $\bi{B}$ at $T_{\mathrm{max}}$ and cooling the system down to $T_{\mathrm{min}}$ at a given temperature sweep rate {$R$} measured in Kelvin per Monte Carlo step (MCS). ZFC runs, on their turn, started with a similar cooling procedure but with $\bi{B}=0$. Once $T_{\mathrm{min}}$ was reached, the desired $\bi{B}$ was applied and the system was heated at the same rate {$R$}.
As mentioned above, MCS were converted to real time within the DSIM by the equivalence $1\,\mathrm{ MCS}\equiv 2.5\,\mathrm{ ms}$~\cite{jaubert2011magnetic}. 
With the intention to compare {\change out-of-equilibrium quantities} obtained at rate $R$ with {\change their} value close to equilibrium, we also performed very slow runs (with a rate $R/10$). Although for the DSIM the slow runs were not in true equilibrium below $\approx 0.6\,\mathrm{K}$ ---for example, no first order transition into the DSIM ground state~\cite{melko2001long,Borzi13} was observed near $0.18\,\mathrm{K}$--- we still refer to them as ``equilibrium'' in the text or figures. {\change Finally, we introduce the monopole density, calculated as 
\begin{equation}
    \rho=\overline{\frac{1}{N_{\mathrm{tetra}}}\sum_{\nu} a_{\nu}} \, , 
\end{equation}
where the Greek index $\nu$ denotes a sum carried over the {diamond}-lattice sites corresponding to the centres of the tetrahedra (the total number of which is $N_{\mathrm{tetra}}$), $a_{\nu}$ is equal to $1$ if there is a monopole in site $\nu$ and equal to $0$ otherwise, and the overline indicates an average over independent runs.}

Experiments were performed on \DTO~single crystals grown by the floating zone method. The magnetisation sample was cut as a thin platelet with its longest side along $[100]$. ZFC and FC magnetization measurements were performed between $T_{\mathrm{min}}$ and $T_{\mathrm{max}}$ in a single crystal oriented with the magnetic field parallel to the crystallographic $[100]$ direction. We used a bespoke magnetometer~\cite{slobinsky2012fast} in a commercial dilution fridge. The protocols were identical to those used in the simulations except for a delay of approximately five minutes {before} the ramp of the magnetic field was started at $T_{\mathrm{min}}$ in ZFC runs, in which no data was acquired.
In order to thermally equilibrate the sample, we coated the biggest sides of the platelet with silver paint, and thermally linked them to the mixing chamber using \chem{Au} and \chem{Cu} wires. {Magnetisation measurements were complemented with ac-susceptibility {experiments} performed in a bespoke probe in a commercial \chem{He^3} cryostat at a fixed frequency $f=1.7\,\mathrm{Hz}$. In all runs the field was applied at $1.7\,\mathrm{K}$ and the susceptibility was measured as a function of temperature with a cooling rate $R=13\,\mathrm{mK}/\mathrm{min}$. The susceptibility probe was immersed in the \chem{He^3} chamber in order to guarantee a good thermal contact.} {We cut samples for different field orientations, aligning the longer side with the magnetic field in order to decrease demagnetisation effects. Sample size was $l = 3.5\,\mathrm{mm}$ in longitude with an area $A = 0.69\times 0.48\,\mathrm{mm^2}$ for the $[111]$ sample, and $l = 4.55\,\mathrm{mm}$ with $A = 0.71 \times 0.66\,\mathrm{mm^2}$ for $[100]$. The magnetic field $B$ was subjected to demagnetisation corrections in order to obtain the internal magnetic field {\change at the blocking temperature $T_{\mathrm{B}}$: $B_{\mathrm{local}}=B- DM(T_{\mathrm{B}})$. $D$ was estimated with standard methods, assuming the samples were perfect, rectangular prisms~\cite{aharoni1998demagnetizing}. In the case of ac-susceptibility, the magnetization at the blocking temperature was approximated with the aid of Monte Carlo simulations, assuming that the sample was then near equilibrium; the correction was of the order of $30\%$ of the applied field.}}


\section{Results and discussion}

\subsection{In-field dynamics with nearest-neighbour interactions}

\begin{figure}
    \centering
    \includegraphics[width=0.7\linewidth]{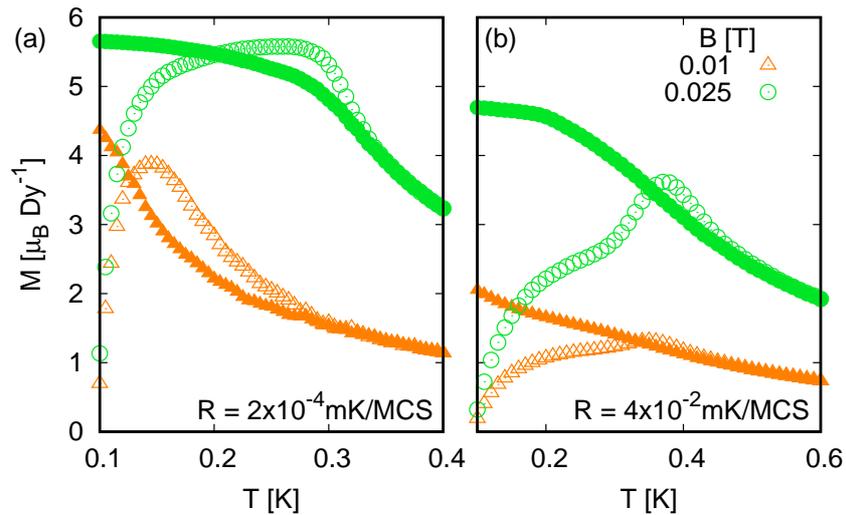}
    \caption{Magnetisation as a function of temperature in zero field cooling (ZFC, open symbols) and field cooling (FC, full symbols) protocols, for the nearest neighbours model (NNSIM). These curves were obtained for two values of the field $\bi{B}\parallel[100]$, and two sweep rates: (a)~{$R=2\times10^{-4}\,\mathrm{mK/MCS}$} and (b)~{$R=4\times10^{-2}\,\mathrm{mK/MCS}$}. In all cases there is a range of temperature where $M^{\mathrm{(ZFC)}}>M^{\mathrm{(FC)}}$.}
    \label{fig:rates_ZFCFC}
\end{figure}

We begin our study with simulations for the NNSIM, concentrating on the magnetisation in an applied field $\bi{B}$ as a function of temperature. Differences between the curves measured using ZFC and FC protocols are the usual signature for departures from equilibrium.
\Fref{fig:rates_ZFCFC} shows the magnetisation as a function of temperature for two sweep rates and under two different values of the magnetic fields along $[100]$. Differences between $M^{\mathrm{(ZFC)}}$ and $M^{\mathrm{(FC)}}$, together with a maximum in $M^{\mathrm{(ZFC)}}$, are only observed below $T\approx 450\,\mathrm{mK}$. This is remarkable taking into account the extremely fast sweep rates (the whole run for $R=4 \times 10^{-2}\,\mathrm{mK/MCS}$ took only $12500\,\mathrm{MCS}$). It is clear that the NNSIM with single-spin-flip dynamics fails to reproduce the most salient feature of spin-ice dynamics in \DTO: its abrupt freezing below $T\approx 650\,\mathrm{mK}$~\cite{snyder2001spin}.
In spite of this evident deficiency, the absence of monopole interactions in the NNSIM makes more apparent the influence of internal constraints in spin dynamics in spin ices, which are at the heart of some of the peculiarities observed in \fref{fig:rates_ZFCFC}. It will be useful to review them, as an advance for the dipolar model results presented in the next section. 

Both sets of curves in \fref{fig:rates_ZFCFC} (with a sweep rate differing by a factor 200) show a range of $T$ in which the ZFC magnetisation is greater than its FC counterpart. We have observed this very unusual effect\footnote{A similar phenomenon was found in manganites~\cite{zhao2006magnetization}, but its explanation --in terms of inhomogeneities and phase coexistence-- does not appear to bear any relation with the physics of spin ice.} with independence of the applied field direction.
Another interesting feature of these curves, germane to the previous one, is the extremely steep growth of $M^{\mathrm{(ZFC)}}$ at very low temperatures even for very small fields. 

The presence of a magnetic field in a given direction favours a subset of the possible spin configurations in a tetrahedron. If the field is turned on at low temperatures ($k_{\mathrm{B}} T_{\mathrm{min}} \lesssim \mu B$ or, in the language of monopoles, $k_{\mathrm{B}} T_{\mathrm{min}} \lesssim a_{\mathrm{dia}}QB$) after a ZFC, massive spin-flipping within the two-in/two-out spin-ice configuration will be required in order to reach the new equilibrium state. {As shown by Castelnovo and collaborators~\cite{Castelnovo2008}, spin flips are equivalent to the creation, annihilation, or movement of preexisting monopoles. At \textit{very low} temperatures, a magnetic field would push any of these monopoles ---increasing thus the magnetisation--- like real charges are pushed by an electric field. They would then travel largely undeflected by thermal noise and unstopped until two of opposite sign meet by chance (annihilation) or all the available paths for their movement have been used (saturation of the magnetisation)~\cite{jaubert2011magnetic}. In this simplistic view the topological nature of these excitations can lead to a magnetisation $M^{\mathrm{(ZFC)}}(T)$ above its equilibrium value at low $T$, joining the equilibrated curve at higher temperatures (and the FC one) from above. However, this argument requires the existence of a finite density of monopoles at very low temperatures, and $\rho$ is expected to be \textit{exponentially small} at low temperatures ($T \ll J_{\mathrm{eff}}$). Thus, sharp variations of the magnetisation are not expected in this regime of temperatures and small fields. The very fast change in $M^{\mathrm{(ZFC)}}$ observed well below $0.2\,\mathrm{K}$ in \fref{fig:rates_ZFCFC} deserves then our special attention. }

{We believe the root of this unexpected behaviour may be connected to a magnetic version of the \textit{second Wien effect} for electrolytes~\cite{Onsager1934deviations,Kaiser2013onsager}, first mentioned in the context of frustrated magnetism in~\cite{bramwell2009measurement} and~\cite{giblin2011creation}. The applied field $B$ pushes apart the \textit{almost} random walks of the opposite charges of a monopole pair; this disfavours their annihilation and thus can increase their concentration~\cite{bramwell2009measurement}. In turn, the enhanced ``carrier'' concentration in the presence of a field would favour a fast change in the magnetisation. Of course, Coulomb interactions between charges are a key ingredient of the Wien effect~\cite{bramwell2009measurement,Onsager1934deviations}. They are present in electrolytes, real spin ices, and the DSIM; however, they are absent within the NNSIM, but for a weak entropic attraction between opposite charges~\cite{henley2010coulomb,ryzhkin2005magnetic,castelnovo2012spin}. It is possible that this explains that the enhanced monopole density could be noticeable within the NNSIM even at very small fields, provided that the temperature is so low that the tendency of the monopoles to diffuse is much smaller than the dragging force of the field ({$k_{\mathrm{B}}T\ll QBa_{\mathrm{dia}}$}~\cite{giblin2011creation}).
There is also another remarkable characteristic of this process which has no counterpart in the electric case, but which should also be present within the DSIM: in spin ices the degree of polarisation of the vacuum from/in which the monopoles are created, live, or die, can change the rate at which they are created/destroyed, and the way in which they can move. A magnetised background, for example, can lead to a current of monopoles even in the absence of a magnetic field~\cite{ryzhkin2005magnetic}.
As we will see, this can explain in the monopole language the marked difference in shape between ZFC and FC curves at the lowest temperatures. }

\begin{figure}
    \centering
    \includegraphics[width=0.7\linewidth]{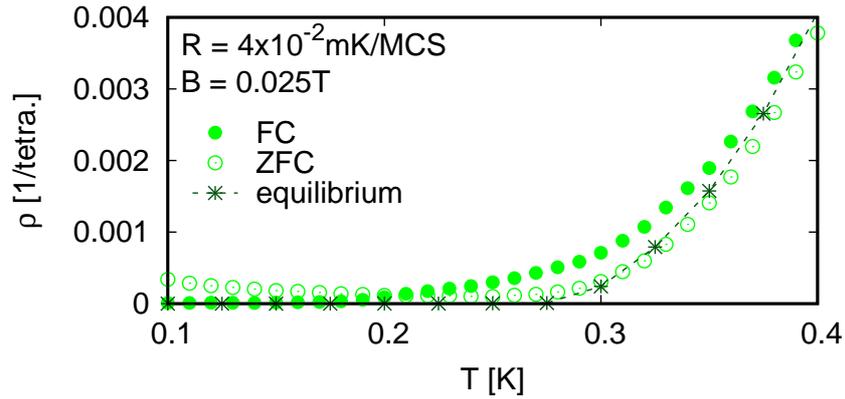}
    \caption{Monopole density {\change in monopoles per tetrahedron (1/tetra.)} measured after ZFC or FC protocols in the NNSIM; the magnetic field used was $B=0.025\,\mathrm{ T}$ along the $[100]$ direction, and the sweep rate {$R=4\times10^{-2}\,\mathrm{mK/MCS}$}. The equilibrium density is also shown for comparison. In spite of the smallness of the field relative to the exchange constant ($\mu B \ll J_{\mathrm{eff}}$), $B$ has a relatively big effect at very low temperatures, with $\rho^{\mathrm{(ZFC)}}>\rho^{\mathrm{(FC)}}$. At high temperatures ($T \gtrsim 0.4\,{\mathrm{K}}$) the three curves have a similar behaviour; the FC (ZFC) density is somewhat bigger (smaller) than the equilibrium curve, accounting for the fact that the system keeps some memory of having being at higher (lower) temperatures, where the monopole density was bigger (smaller).} 
    \label{fig:rho_ZFCFC}
\end{figure}

\Fref{fig:rho_ZFCFC} shows the behaviour of the monopole density $\rho$ as a function of temperature for the FC and ZFC protocols for $B=0.025\,\mathrm{T}$ along $[100]$ and sweep rate $R=4\times10^{-2}\,\mathrm{mK/MCS}$. The curves correspond to the green ones in \fref{fig:rates_ZFCFC}(b); {the monopole density in equilibrium at the same field is also included for comparison}. 
The magnetic field is very small relative to the exchange energy\footnote{In order to give an idea of the order of magnitude we can remember that for $\bi{B}\parallel[111]$ single monopoles are stable only for $B$ of the order of $1\,\mathrm{T}$ ($40$ times bigger).}, but evidently has a real effect in $\rho$. 
At the lowest temperatures, where the dragging force of the field relative to thermal diffusion should be at its peak, the behaviour of the ZFC and FC curves diverge; the differences are rather small in absolute terms, but huge in relative ones: $\rho^{\mathrm{(ZFC)}}/\rho^{\mathrm{(FC)}} \sim  40$ at $T_{\mathrm{min}}$. The out-of-equilibrium monopole concentration $\rho^{\mathrm{(ZFC)}}$ decays as $T$ increases, and reaches a minimum when $T \approx 0.25\,\mathrm{K}$. At higher $T$ it follows the equilibrium curve, rising due to the increasing thermal fluctuations which also diminish the effectiveness of the dragging force of the field. It is clear that $\rmd M^{\mathrm{(ZFC)}}/\rmd T$ mimics this behaviour, something that will be explored further in the next section for the more relevant case of the DSIM. We simply note here that, when no monopole interactions are present, the out-of-equilibrium monopole current dragged by the magnetic field $B$ at low temperatures can be so big that its cumulative effect is to increase $M^{\mathrm{(ZFC)}}$ over the FC value, so that it joins the equilibrium curve from \textit{above}.

\begin{figure}
    \centering
    \includegraphics[width=0.7\linewidth]{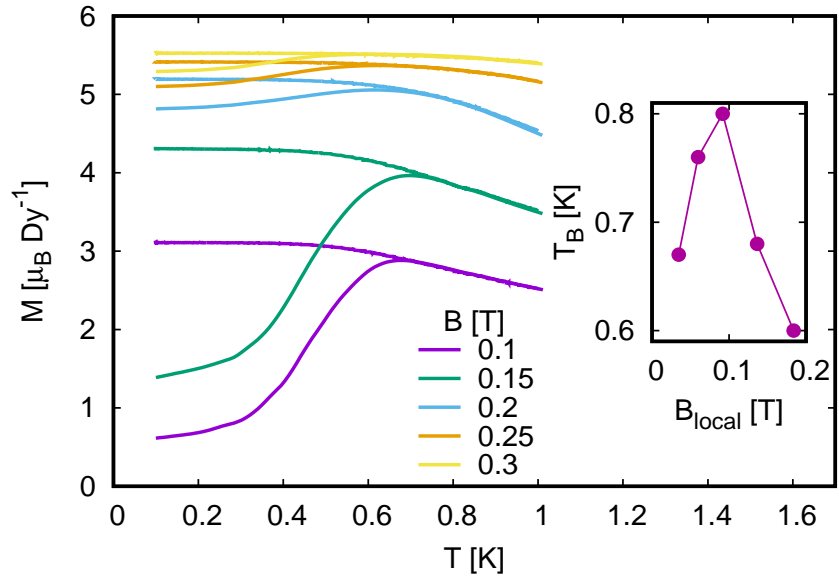}
    \caption{Experimental ZFC-FC magnetisation measurements in \DTO\ single crystals with field applied along the $[100]$ direction and a sweep rate $R=5\,\mathrm{mK}/\mathrm{min}$. Due to the density of points we chose to use lines instead of symbols (ZFC curves always lie below or coincide within errors with their respective FC ones). The inset shows the blocking temperature as a function of the local magnetic field $B_{\mathrm{local}}$, calculated after subtracting the demagnetising field. ${\change T_{\mathrm{B}}^{\mathrm{mag}}}$ does not evolve monotonically with the intensity of the field inside the material, with a marked peak near $B_{\mathrm{local}} \approx 0.1\,\mathrm{T}$.}
    \label{fig:demian}
\end{figure}

{While the ZFC magnetisation curve over the FC one is very interesting from a theoretical viewpoint, it has not been experimentally observed either in powders or single crystals (see~\cite{snyder2004low} and~\fref{fig:demian}). This, together with other obvious shortcomings of the NNSIM mentioned before, makes us turn to the DSIM. Before doing so we want to stress how remarkable it is that despite the simplicity of the NNSIM, there are some features in the ZFC-FC curves in \fref{fig:rates_ZFCFC} that do have a correspondence in previous experiments made on \DTO\ powders~\cite{snyder2004low} which have not been studied theoretically/numerically in the past: \textit{i-}~The experiments also show the rapid increase of the ZFC magnetisation curves at the lowest temperatures; different from our previous simulations, they only appear above a certain threshold field, of the order of $0.2\,\mathrm{T}$.
\textit{ii-}~The temperature at which the ZFC curves attains a maximum \textit{increases} with increasing field (\fref{fig:rates_ZFCFC}). This ---as stressed by Snyder and collaborators~\cite{snyder2004low}--- goes against the usual trend found in systems with slow dynamics~\cite{de1978stability,lefloch1994spin,fukaya2001magnetic}. As we will discuss in the next section, this feature is also present in our magnetisation measurements (\fref{fig:demian}) performed on single crystals under fields along $[100]$.}


\subsection{In-field dynamics with dipolar interactions and experimental results}

\begin{figure}
    \centering
    \includegraphics[width=0.7\linewidth]{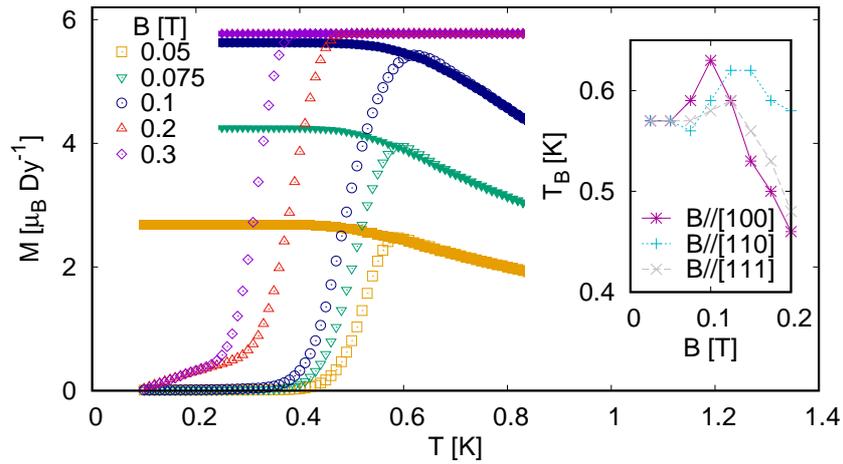}
    \caption{Magnetisation as a function of the temperature in a ZFC (open symbols) - FC (full symbols) protocol in the DSIM, for different values of the field along the $[100]$ direction and a sweep rate $R=50\,\mathrm{mK/min}$. The blocking temperature ${\change T_{\mathrm{B}}^{\mathrm{mag}}}(B)$ is independent of the field direction at low fields, but presents a clear peak for $\bi{B}\parallel [100]$ near $B \approx 0.1\,\mathrm{T}$ (inset).}
    \label{fig:ZFCFC_MC}
\end{figure}

Although it does not capture all its complex features~\cite{takatsu2013two,slobinsky2010unconventional,revell2013evidence,takatsu2013ac}, the zero-field dynamics of the DSIM with Metropolis single-spin-flip moves is comparable to that of real spin ice materials~\cite{jaubert2009signature}. This allows us to measure the temperature-sweep rate of our simulations in real units with a certain degree of approximation. Also, the slower dynamics permits exploring magnetic fields and temperature sweep rates within the ranges used in experimental measurements without reaching saturation. \Fref{fig:ZFCFC_MC} shows the ZFC-FC magnetisation in the DSIM for different values of $B$ along the $[100]$ direction in the DSIM at a sweep rate {$R=2\times10^{-3}\,\mathrm{mK/MCS}$} or, following~\cite{jaubert2011magnetic}, $R \approx 50\,\mathrm{mK/min}$. 
Differently from the NNSIM, and in accordance with experiments (see \fref{fig:demian} and~\cite{snyder2004low}), we note that now the simulated ZFC magnetisation curves tend to remain (but for a very narrow $T$ interval near their maximum, {\change which does not seem to diminish with system size}) below their corresponding FC ones. {The characteristic temperatures signaling out-of-equilibrium behaviour are higher than those for the NNSIM, but still (in spite of the faster sweep rate) somewhat low compared with experiments (\fref{fig:demian}).

At the lowest temperatures, the ZFC curves} are very flat (near $M^{\mathrm{(ZFC)}}=0$) for low $B$. Only when the magnetic field reaches $B_{\mathrm{th}} \approx 0.2\,\mathrm{T}$ there is a finite slope at $T_{\mathrm{min}}$. {Although it is quite significant, the initial {\change increase} in $M^{\mathrm{(ZFC)}}$ for fields above this threshold} seems to be much less pronounced than in the experimental curves (see figure~1 in~\cite{snyder2004low}, and our \fref{fig:demian} above\footnote{{Due to the measurement protocol implemented, the initial {\change increase} of $M^{\mathrm{(ZFC)}}$ has not been recorded. However, the overall result of it can be read as the value of $M^{\mathrm{(ZFC)}}(T_{\mathrm{min}})$ in \fref{fig:demian}.}}). {This difference may be related to the release of heat from the magnetic to the elastic degrees of freedom in real samples, which can lead to magnetic deflagration~\cite{slobinsky2010unconventional}. We clearly have not considered this type of magnetoelastic coupling in our simulations, but the spin dynamics after small avalanches triggered by low fields have been recently studied in detail in~\cite{paulsen2014far} and~\cite{paulsen2016experimental}}. 

In spite of the aforementioned differences, the threshold field $B_{\mathrm{th}}$ needed to awake fast dynamics at $100\,\mathrm{mK}$ in the ZFC curves is in very good coincidence in experiments and numerical results. A simple calculation shows that the field $B$ at which the magnetic push $BQ$ over a monopole of charge $Q$ is equal to the pull due to a charge $-Q$ at a distance $a_{\mathrm{dia}}$ is $B \approx 0.24\,\mathrm{T}$. We thus find that $B_{\mathrm{th}}$ is not far from the magnetic field needed to transform most bound pairs of $+$ and $-$ monopoles into free charges near $T=0$~\cite{giblin2011creation}.

\begin{figure}
    \centering
    \includegraphics[width=0.7\linewidth]{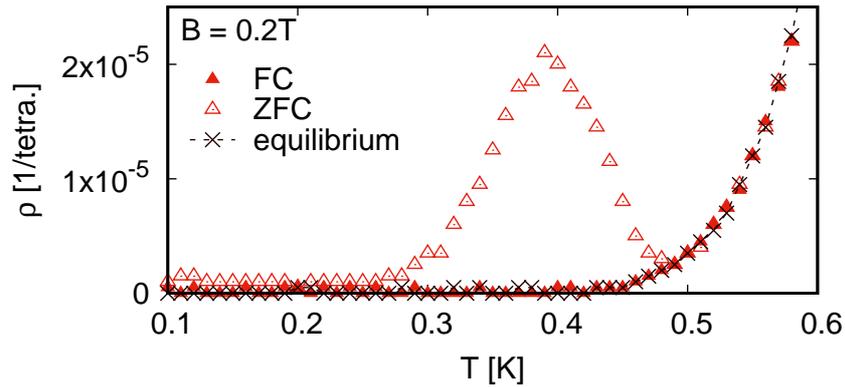}
    \caption{Density of magnetic monopoles after ZFC or FC protocols within the DSIM; the magnetic field $B=0.2\,\mathrm{T}$ was applied along the $[100]$ direction and the sweep rate was $R=50\,\mathrm{mK/min}$. We note again that $\rho^{\mathrm{(ZFC)}}$ appears enhanced with respect to $\rho^{\mathrm{(FC)}}$ at the lowest temperatures, peaking near $T \approx 0.4\,\mathrm{K}$. The density at equilibrium is shown as a reference.}
    \label{fig:rho_dip}
\end{figure}

\Fref{fig:rho_dip} shows the monopole density as a function of $T$ for the different protocols in the presence of a field $B=0.2\,\mathrm{T}$. It is analogous to \fref{fig:rho_ZFCFC}, now simulated with the DSIM. As happened in the NNSIM case, $\rho^{(ZFC)}(T)$ appears enhanced at low temperatures, in apparent correlation with the finite initial slope in $M^{\mathrm{(ZFC)}}(T)$. The ratio $\rho^{\mathrm{(ZFC)}}/\rho^{\mathrm{(FC)}}$ reaches a maximum above $40$ near $T = 0.4\,\mathrm{K}$. After this peak, $\rho^{\mathrm{(ZFC)}}(T)$ follows a pattern similar to the NNSIM: it reaches a minimum and grows due to thermal excitation near the equilibrium values. On its turn, the FC curve follows (within errors) the equilibrium curve in the whole temperature range. 

{The conditions under which our magnetisation measurements and simulations were performed imply a regime very different to that in the first studies on the Wien effect in spin ices, made in~\cite{bramwell2009measurement}. Although the temperature range is similar, our magnetic fields are typically three orders of magnitude bigger than in~\cite{giblin2011creation}, and $10$ times those used in \fref{fig:rates_ZFCFC}. On the other hand, the studies on non-linear behaviour in~\cite{kaiser2015ac} and~\cite{paulsen2016experimental} were performed under conditions similar to ours (standard for $M$ vs. $T$ curves at very low temperatures).
The features we observe in the magnetisation and density of monopoles at low temperatures strongly suggest a connection with Wien dissociation. However, adding more complexity to this, the varying temperature, and the influence on the dynamics of an energy balance perturbed by a significant applied field, also play an important role in shaping our $M(T)$ curves.
} 

In order to try a rough quantitative analysis of these curves at low temperatures, we note that in this regime very far from equilibrium ({turning {\change on} a field after a ZFC to $T=0.1\,\mathrm{K}$, with a sweep rate which is exceedingly fast for the equilibration times at these temperatures}) we can expect temperature changes to be similar to time derivatives: $\rmd M^{\mathrm{(ZFC)}}/\rmd T \approx (1/R)\, \rmd M^{\mathrm{(ZFC)}}/\rmd t$, with $t$ the Monte Carlo time. This fact, together with the aforementioned likeness between the density of monopoles $\rho^{\mathrm{(ZFC)}}(B,T)$ curve (\fref{fig:rho_dip}) and the slope $\rmd M^{\mathrm{(ZFC)}}/\rmd T$ (\fref{fig:ZFCFC_MC}) make us think that we are in a condition in which the biggest contribution to magnetisation changes comes from a current of monopoles being dragged by the magnetic field. {Since we are dealing with the ZFC case, we will neglect the entropic drive proportional to the magnetisation~\cite{ryzhkin2005magnetic}.} In analogy with Ohm's law we then propose:
\begin{equation}\label{dMdT}
    \frac{\rmd M^{\mathrm{(ZFC)}}(B,T)}{\rmd T} =  \tilde{m} \rho^{\mathrm{(ZFC)}}(B,T)  B \, . 
\end{equation}
The factor $\tilde{m}$ seems to be analogous to the mobility in semiconductors or electrolytes, but note that it multiplies the total density of monopoles ---not just that of free ones.
The use of $\rho^{\mathrm{(ZFC)}}(B,T)$ in this equation allows for thermal effects (i.e., to consider some of the temperature evolution, and not \textit{just} the evolution in time with other units). However, this effect is limited since $\tilde{m}$ is taken simply as a multiplicative constant, independent of \textit{both} temperature and field. To our knowledge, there is no  \textit{a priori} reason why transport in this field and temperature regime should be Ohmic-like\footnote{Note that this equation is not truly linear in $B$, since the magnetic field also enters through the monopole density $\rho(B,T)$.}~\cite{giblin2011creation}. However, this is the simplest attempt to describe the magnetisation curves at low temperature taking into account the observed facts. 

\begin{figure}
    \centering
    \includegraphics[width=0.7\linewidth]{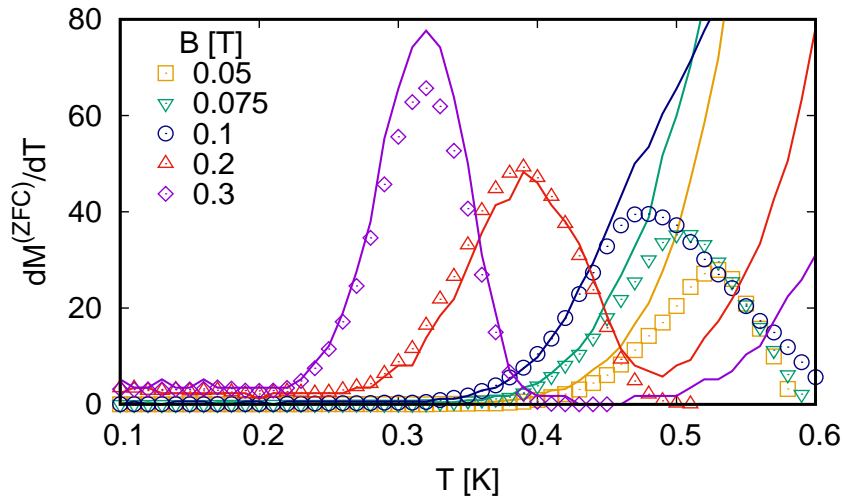}
    \caption{Slope of the ZFC magnetisation $\rmd M^{\mathrm{(ZFC)}}/\rmd T$ as a function of temperature for different values of the field along the $[100]$ direction and a sweep rate $R=50\,\mathrm{mK/min}$. We compare it with a magnetic equivalent of Ohm's law (equation~\eref{dMdT}), plotted as lines with the same colour code. Both sets of curves are comparable within the regime given by the condition $\gamma = 0.075\,\mathrm{T\,K^{-1}} \times T/B \ll 1$.}
    \label{fig:dMZFCdT_lowT}
\end{figure}

\Fref{fig:dMZFCdT_lowT} compares both sides of~\eref{dMdT} for $\tilde{m}=1.15\times10^7\, \mu_{\mathrm{B}}\,{\mathrm{Dy^{-1}\,T^{-1}\,K^{-1}}}$,
with $\rho$ in units of monopoles per tetrahedron. Assuming that for $B=0.24\,\mathrm{T}$ all monopoles are free, this translates into a speed of $\approx 1$ monopole move per MCS. {This is very near to the value of $1.5$ moves per MCS, expected for a monopole dragged by a magnetic field along $[100]$ at the lowest temperatures in an unpolarised ``two-in/two-out'' background.\footnote{{The value of 1.5 moves per MCS (or every $\approx 2.5\,\mathrm{ms}$) is the speed limit in these conditions~\cite{Kaiser2013onsager}.}}} Going back to \fref{fig:dMZFCdT_lowT}, we first note that in the lowest temperature regime where the field drag dominates over diffusion {($\gamma\equiv k_{\mathrm{B}}T/ QBa_{\mathrm{dia}} \ll 1$)}
our single-parameter equation accounts well for the initial slope, which is fuelled by the enhanced $\rho^{\mathrm{(ZFC)}}$. At high fields ($B \geq 0.2\,\mathrm{T}$) the expression can even explain reasonably well the peak in the magnetisation slope (by means of its twin peak in the monopole density, as exemplified in \fref{fig:rho_ZFCFC}). This is noteworthy, considering that the mobility is likely to be reduced when the system (the vacuum of monopoles) increases its polarisation, due to the entropic pull.
On the other hand, at lower fields $\rmd M^{\mathrm{(ZFC)}}/\rmd T$ peaks at $T$ such that $\gamma \lesssim 1$, where our expression fails. Finally,~\eref{dMdT} can only account for the curve at the smallest fields at the lowest temperatures.
The relative contribution of bound pairs of monopoles to $\rmd M^{\mathrm{(ZFC)}}/\rmd T$ is expected to decrease as the field increases. This justifies its exclusion from~\eref{dMdT} in the high-field regime of our measurements, and could be in part responsible for the observed failure of this equation at low fields. Given the simplicity of~\eref{dMdT}, we think that its success to describe ZFC magnetisation curves is quite remarkable. More so if we take into account that the range of fields and temperatures where it works better encompasses four regions with very different regimes (true Ohmic regime, pair unbinding, ideal Wien effect, breakdown regime), as illustrated in figure 4 of~\cite{paulsen2016experimental}.

{\subsection{Evolution of the blocking temperature with magnetic field}

After the previous studies at the lowest temperatures, we concentrate now on the intermediate range of $T$ where the onset of the dynamical freezing takes place.} The simulated curves for the DSIM in \fref{fig:ZFCFC_MC} are comparable to the experiments\footnote{{While these curves are much faster than the experimental ones, we have checked than the same physics holds for rates five times slower (i.e., approximately twice the one used in experiments).}} reported in~\cite{snyder2004low} and to our measurements in single crystals for $\bi{B}\parallel[100]$ shown in \fref{fig:demian}. We will use them to study the blocking temperature of the system. Different criteria exist for the determination of  ${\change T_{\mathrm{B}}^{\mathrm{mag}}}(\bi{B})$: the position of the maximum of $M^{\mathrm{(ZFC)}}$~\cite{fiorani1999collective,papusoi1997initial}, the point where the ZFC-FC curves separate, or the temperature at which $M^{\mathrm{(ZFC)}}$ moves away from a Curie's law-like behaviour~\cite{chamberlin1982h}. {\change Due to its straightforward application to numerical data, we took advantage of the first criterion in those fields where low-temperature saturation was not reached. For the other curves, where $M^{\mathrm{(ZFC)}}$ was maximum (and equal to its saturation value) in a range of $T$, we defined $T_{\mathrm{B}}^{\mathrm{mag}}$ as the lower bound of that range.} 
{We used these data to put together the inset to \fref{fig:ZFCFC_MC}, showing the dependence of the blocking temperature ${\change T_{\mathrm{B}}^{\mathrm{mag}}}$ with the field intensity for the three directions of interest. At low fields, an increasing trend of ${\change T_{\mathrm{B}}^{\mathrm{mag}}}(B)$ with field (contrary to most slow dynamics systems) is most clearly observed for $\bi{B}\parallel [100]$. In consequence, ${\change T_{\mathrm{B}}^{\mathrm{mag}}}(B)$ describes a peak for this field direction, which is not as evident for $[110]$ and $[111]$. }

Notably, our ZFC-FC magnetisation measurements on a \DTO\ single crystal under fields along $[100]$ at a rate $R=5\,\mathrm{mK}/\mathrm{min}$ (\fref{fig:demian}) present the same trend.\footnote{{We note that ${\change T_{\mathrm{B}}^{\mathrm{mag}}}$ is higher for the simulations, in spite of the fact that $R$ is (nominally) $10$ times faster than the one in the experiments. This is suggesting ---as happens with thermodynamics~\cite{borzi2016intermediate,Henelius2016refrustration,yavors2008,grigera2015intermediate}--- a limitation of the DSIM in its description of \DTO\ dynamics}.} In fact, ${\change T_{\mathrm{B}}^{\mathrm{mag}}}$ has a similar dependence on the internal field as the DSIM simulations, with a peak near $0.1\,\mathrm{ T}$ (\fref{fig:demian}, inset). 
Since the same effect was observed by Snyder \textit{et al.}~\cite{snyder2004low} in polycrystalline samples of \DTO, we conjecture that it might be due to the contribution of the $[100]$ direction above all others. 
{Both experiments and simulations show an abrupt change in the behaviour of ${\change T_{\mathrm{B}}^{\mathrm{mag}}}(B)$ at fields bigger than $0.1\,\mathrm{T}$ (insets to \fref{fig:demian} and~\ref{fig:ZFCFC_MC}), where the blocking temperature recovers its usual decreasing trend. We believe that the coincidence of the ZFC and FC magnetisation curves at lower temperatures on increasing fields is not truly related with faster dynamics (shorter correlation times), but due mainly to the closeness to saturation for both curves. This will be confirmed below by other experiments.}

\begin{figure}
    \centering
    \includegraphics[width=0.7\linewidth]{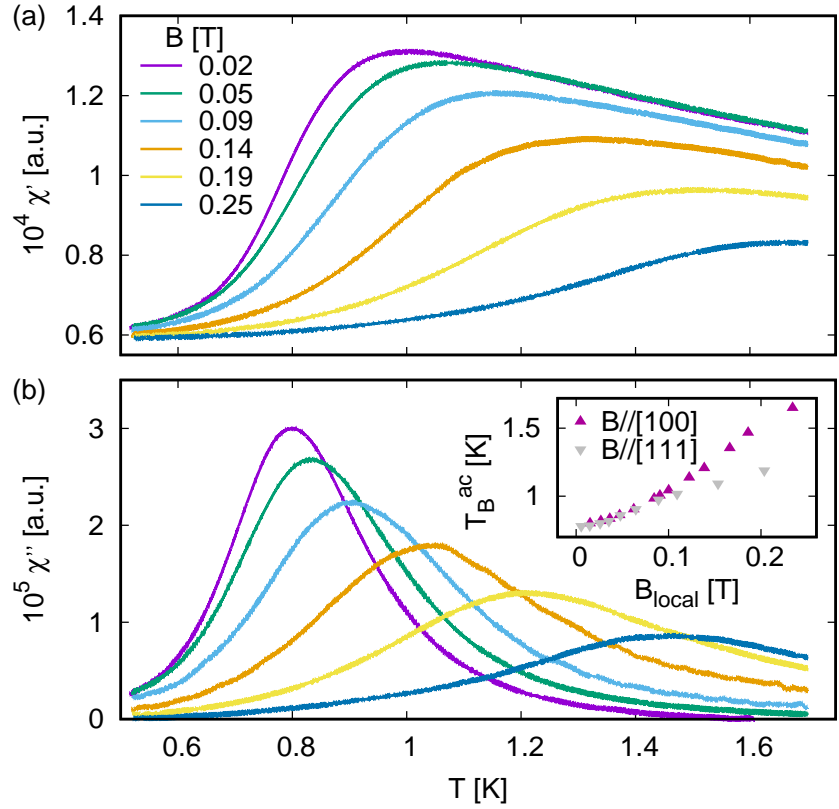}
    \caption{(a)~Real and (b)~imaginary part of the magnetic ac-susceptibility measured in \DTO\ single crystals with a frequency $f=1.7\,\mathrm{ Hz}$ and different values of the dc field applied along the $[100]$ direction. The behaviour of the blocking temperature defined as the position of the maximum of $\chi''$ is independent of the direction of the field only for small {\change fields} (inset). As observed in the simulations, dynamical freezing occurs at higher temperatures for $\bi{B}\parallel[100]$.}
    \label{fig:lucas}
\end{figure}

We have further studied the dependence of the blocking temperature with $\bi{B}$ oriented along the single-crystal directions $[100]$ and $[111]$ using ac-susceptibility measurements. \Fref{fig:lucas} shows the real and imaginary parts of the dynamic susceptibility ($\chi'$ and $\chi''$, respectively) for $\bi{B}\parallel[100]$ at a frequency $f=1.7\,\mathrm{Hz}$. We have defined the blocking temperature $T_{\mathrm{B}}^{\mathrm{ac}}(B)$ for this technique as the temperature at which $\chi''(T,B)$ has its maximum for the given field, and plotted it for the two different crystal orientations in the inset to \fref{fig:lucas}(b).
Since the characteristic measurement time for magnetisation (of the order of seconds) is longer than the inverse of the frequency used, the blocking temperatures we observe are much higher than those obtained in the ZFC magnetisation case. A more drastic difference is the absence of a peak in the blocking temperature for both field orientations: as was also observed in polycrystals~\cite{snyder2004low}, $T_{\mathrm{B}}^{\mathrm{ac}}$ increases monotonically with field. Differently from magnetisation, even near saturation this technique is able to tell us about the ability of the few remaining monopoles to oscillate in or out of phase with the ac field. We conclude that the dynamics continues to slow down (for all field directions) with increasing field. 

The coincidences between the diverse experimental techniques and also simulations seem to be much more important than their differences. Ac susceptibility in single crystals confirms the anomalous increase in $T_{\mathrm{B}}$ observed in $M^{\mathrm{(ZFC)}}$ with increasing field. While at very low fields the behaviour seems to coincide for both field directions, above $\sim 0.05-0.08\,\mathrm{T}$ the increasing trend seems to be reinforced for fields along $[100]$ (insets to~\fref{fig:demian},~\ref{fig:ZFCFC_MC}, and~\ref{fig:lucas}). 
Below we suggest two (not unrelated) possible reasons for the enhanced $T_{\mathrm{B}}(\bi{B})$ when $\bi{B}\parallel[100]$.

\textit{i- Monopole density.} {In the intermediate range of temperatures in which the blocking occurs (not far from equilibrium) the effect of $B$ is to \textit{reduce} the density of magnetic monopoles. This reduction in $\rho(\bi{B},T)$ is much more drastic for $[100]$ than for the other field directions. \Fref{fig:rho_eq} shows nine curves for $\rho(\bi{B},T)$ in equilibrium at low temperatures for three moderate fields along the three relevant directions. It shows the extreme suppression of monopoles (and thus of spin dynamics) for the $[100]$ direction compared with $\bi{B}$ parallel to $[110]$ and $[111]$ for fields above $\approx 0.1\,\mathrm{T}$. It will be easier to illustrate the mechanism taking the $M^{\mathrm{(FC)}}$ case as an example, coming from a state of equilibrium at $T$ (assumed low) and decreasing the temperature to $T-\Delta T$. At the given $B$ and $T$ there will be less monopoles for $\bi{B}\parallel[100]$ than for the other directions (\fref{fig:rho_eq}), and thus less possible spin flips in the characteristic measurement time. On decreasing the temperature to $T-\Delta T$ (and thus decreasing even more these densities and the rates of accepted flips, see \fref{fig:rho_eq}) it is thus to be expected for $M^{\mathrm{(FC)}}(T)$ to fall out of equilibrium first for a field along $[100]$ than for the other field directions. {In more general terms, the dependence of the dynamical arrest on $\bi{B}$ can be thought of as reflecting that of the characteristic relaxation time $\tau$ on monopole density, predicted to be $\tau \propto 1/\rho$ at zero field~\cite{castelnovo2011debye,ryzhkin2005magnetic}. }
 \begin{figure}
     \centering
     \includegraphics[width=0.7\linewidth]{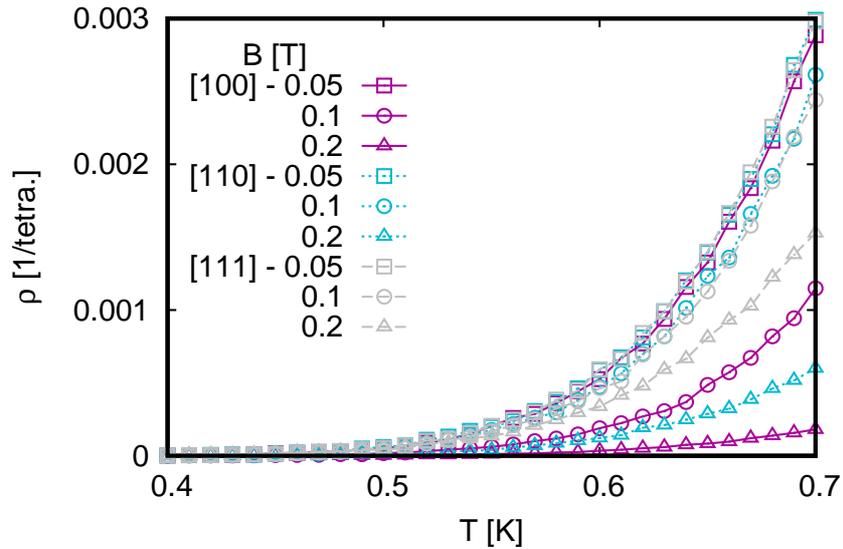}
     \caption{{Equilibrium} monopole density as a function of the temperature in the DSIM for different values of the field in the three directions of interest. Although $\rho$ always decreases with the field in the range of $B$ and $T$ studied, monopole suppression is much more noticeable for $\bi{B}\parallel[100]$.}
     \label{fig:rho_eq}
 \end{figure}

\textit{ii- Kasteleyn transition.} It is interesting to note that a very peculiar, topological transition is expected for $\bi{B}\parallel[100]$, which consists in a sudden saturation of the magnetisation at a \textit{finite} temperature $T_{\mathrm{K}}$ when the concentration of monopoles $\rho$ can be considered negligible~\cite{jaubert2009topological,jaubert2008three}. Within the DSIM, $T_{\mathrm{K}} \approx -0.46\,\textrm{K}+11.16\,\mathrm{K\,T^{-1}}\times B$ at moderate fields~\cite{baez20163d}. This leads to transition temperatures well below the blocking temperature $T_{\mathrm{B}}(B)$ at low fields, but above $1\,\mathrm{K}$ for $B>0.15\,\mathrm{T}$. Due to the non-negligible concentration of monopolar excitations at these temperatures, we would expect this transition to be somewhat rounded and shifted at these temperatures. An intriguing possibility is then that the curve $T_{\mathrm{B}}^{{\mathrm{ac}}}$ for $[100]$ could be explained by two different regimes: a low field regime ---shared with the other field directions--- linked to the usual ($B=0$) freezing; a second regime, for fields such that $T_{\mathrm{K}}(B) > T_{\mathrm{B}}$, related to dynamical arrest associated with the Kasteleyn transition. 
}

\section{Conclusions}

{We have studied some peculiarities of the dynamics in spin-ice models and samples in the presence of a magnetic field $\bi{B}$, using numerical simulations and magnetic measurements in single crystals of \DTO. Firstly, using the nearest-neighbours model we identified indications of physics associated with Wien-like  dissociation in the zero-field cooling (ZFC) magnetisation $M^{\mathrm{(ZFC)}}$ vs. temperature curves. A field applied after cooling the sample at the lowest temperatures enhances the density of monopoles; this leads to a big current of monopoles and a pronounced slope in $M^{\mathrm{(ZFC)}}(T)$. The current can be so big within the nearest-neighbour model that within a range of temperatures the ZFC magnetisation curve lies above the curve measured while cooling with the field applied (FC).
Indeed, the vacuum polarization (an ingredient missing in electrolytes) affects the creation and movement of the monopoles, shaping quite differently the ZFC and FC curves: a modest applied field of $B=0.025\,\mathrm{T}$ changes the density of monopoles by a factor of $\approx 40$. Real materials (polycrystals, and our single crystals) also show the presence of a big slope in the ZFC magnetisation vs temperature curve at very low temperatures. However, due to the interaction between excitations, they do so only after a threshold field of $B_{\mathrm{th}} \approx 0.2\,\mathrm{T}$ is reached, and with the field-cooling magnetisation always above $M^{\mathrm{(ZFC)}}(T)$. The dipolar model confirms this threshold field, which in the monopole picture can be interpreted as the minimum field needed to have a significant number of free $+$ and $-$ monopoles (as opposed to bound $+-$ pairs). 
In the range of fields of our magnetisation vs. temperature measurements ($B \in [0.05,0.3]\,\mathrm{T}$) we propose a version of a magnetic Ohm's law to describe the currents of these free monopoles, and hence the ZFC curves. This approach is valid in the low temperature region $k_{\mathrm{B}}T \ll QBa_{\mathrm{dia}}$ (a sort of ballistic regime, where field drag dominates over monopole diffusion). Within this region, changes in the magnetisation can be directly related to (or used to measure) the monopole density.

One target of our work was to study the evolution of the blocking temperature (the temperature $T_{\mathrm{B}}(\bi{B})$ at which the system falls out of equilibrium within the characteristic timescale of the measurements) with the magnetic field. Measurements on single crystals confirm the results ---discovered previously in powders and contrary to those found in other slow dynamic systems--- of an increasing $T_{\mathrm{B}}$ with field. Quite remarkably, this increasing trend is present even within the nearest-neighbour model. The dipolar model describes well the behaviour of ${\change T_{\mathrm{B}}^{\mathrm{mag}}}(\bi{B})$ {\change (extracted from the magnetisation measurements)}, with a marked peak for $\bi{B}\parallel[100]$ near $0.1\,\mathrm{T}$. As a possible origin of the {\change increase} of the blocking temperature we propose the suppression of monopoles with increasing field. The depletion of monopoles with field is enhanced for fields along $[100]$, something that could explain (together with the closeness to a Kasteleyn transition) the stronger dependence of $T_{\mathrm{B}}$ with field for this crystallographic direction. }

\ack
{We thank P.~Holdsworth for useful comments on our manuscript.} This work was supported by Agencia Nacional de Promoci\'on Cient\'\i fica y Tecnol\'ogica (ANPCyT) through grants PICT 2013-2004, PICT 2014-2618 and PICT 2017-2347, and Consejo Nacional de Investigaciones Cient\'\i ficas y T\'ecnicas (CONICET) through grant PIP 0446.


\section*{References}


\providecommand{\newblock}{}

\end{document}